\documentclass[a4paper,11pt]{article}
\pdfoutput=1 

\usepackage{jinstpub} 

\title{\boldmath Update of the trigger system of the PandaX-II experiment}

\author[a]{Qinyu Wu,}
\author[a]{Xun Chen,}
\author[a,b,c,d]{Xiangdong Ji,}
\author[a]{Jianglai Liu,}
\author[a]{Siao Lei,}     
\author[a]{Xiangxiang Ren,}
\author[e]{Meng Wang,}
\author[a]{Mengjiao Xiao,}
\author[a]{Pengwei Xie,}
\author[a]{Binbin Yan,}
\author[a]{Yong Yang}

\affiliation[a]{INPAC and Department of Physics and Astronomy, Shanghai Jiao Tong University, Shanghai Laboratory for Particle Physics and Cosmology, Shanghai 200240, China}
\affiliation[b]{Center of High Energy Physics, Peking University, Beijing 100871, China}
\affiliation[c]{Department of Physics, University of Maryland, College Park, Maryland 20742, USA}
\affiliation[d]{T. D. Lee Institute, Shanghai 200240, China}
\affiliation[e]{School of Physics and Key Laboratory of Particle Physics and Particle Irradiation (MOE), Shandong University, Jinan 250100, China}

\emailAdd{renxx@sjtu.edu.dn}
\emailAdd{yong.yang@sjtu.edu.dn}

\abstract{PandaX-II experiment is a dark matter direct detection experiment using about half-ton of liquid xenon as the sensitive target. The electrical pulses detected by photomultiplier tubes from scintillation photons of xenon are recorded by waveform digitizers. The data acquisition of Pandax-II relies on a trigger system that generates common trigger signals for all waveform digitizers. Previously an analog device-based trigger system was used for the data acquisition system. In this paper we present a new FPGA-based trigger system. The design of this system and trigger algorithms are described. The performance of this system on real data is presented.} 

\keywords{Dark matter experiment, Digital signal processing, Trigger algorithms.}

\begin{document}
\maketitle
\flushbottom

\section{Introduction of PandaX-II and its data acquisition system}
\label{sec:intro}

PandaX-II experiment is dark matter (DM) direct detection experiment,
located at the China Jin-Ping Underground Laboratory. The primary goal
of PandaX-II experiment is to search for weakly interacting massive
particle (WIMP)-like DM. PandaX-II operates a dual phase time
projection chamber (TPC). The TPC contains 580 kg liquid xenon in the
sensitive volume enclosed by polytetrafluoroethylene (PTFE) reflective
panels with an inner diameter of 646~mm and a vertical maximum drift
length of 600~mm defined by the cathode mesh and gate grid. For each
interaction in liquid xenon, both prompt scintillation photons (S1)
and delayed electroluminescence photons (S2) are collected by two
arrays of 55 Hamamatsu R11410-20 photomultiplier tubes (PMTs) located
at the top and bottom, respectively.  The time difference between S1
and S2 gives the vertical location of the interaction point in liquid xenon
and the charge pattern of S2 signals on the top PMT array gives the
horizontal position. The 3-D interaction position can be used to
reduce ambient background from detector and surrounding materials. In
addition, the ratio of S2 (in unit of Photoelectron,PE) and S1 can be
used to discriminate WIMP signals against $\beta$ and $\gamma$
backgrounds.  More detailed descriptions of the PandaX-II experiment
can be found in Ref.~\cite{PandaX-II_com, PandaX-II_100, PandaX-II_100SD}.

\begin{figure}[h]\centering
  \includegraphics[width=0.95\textwidth]{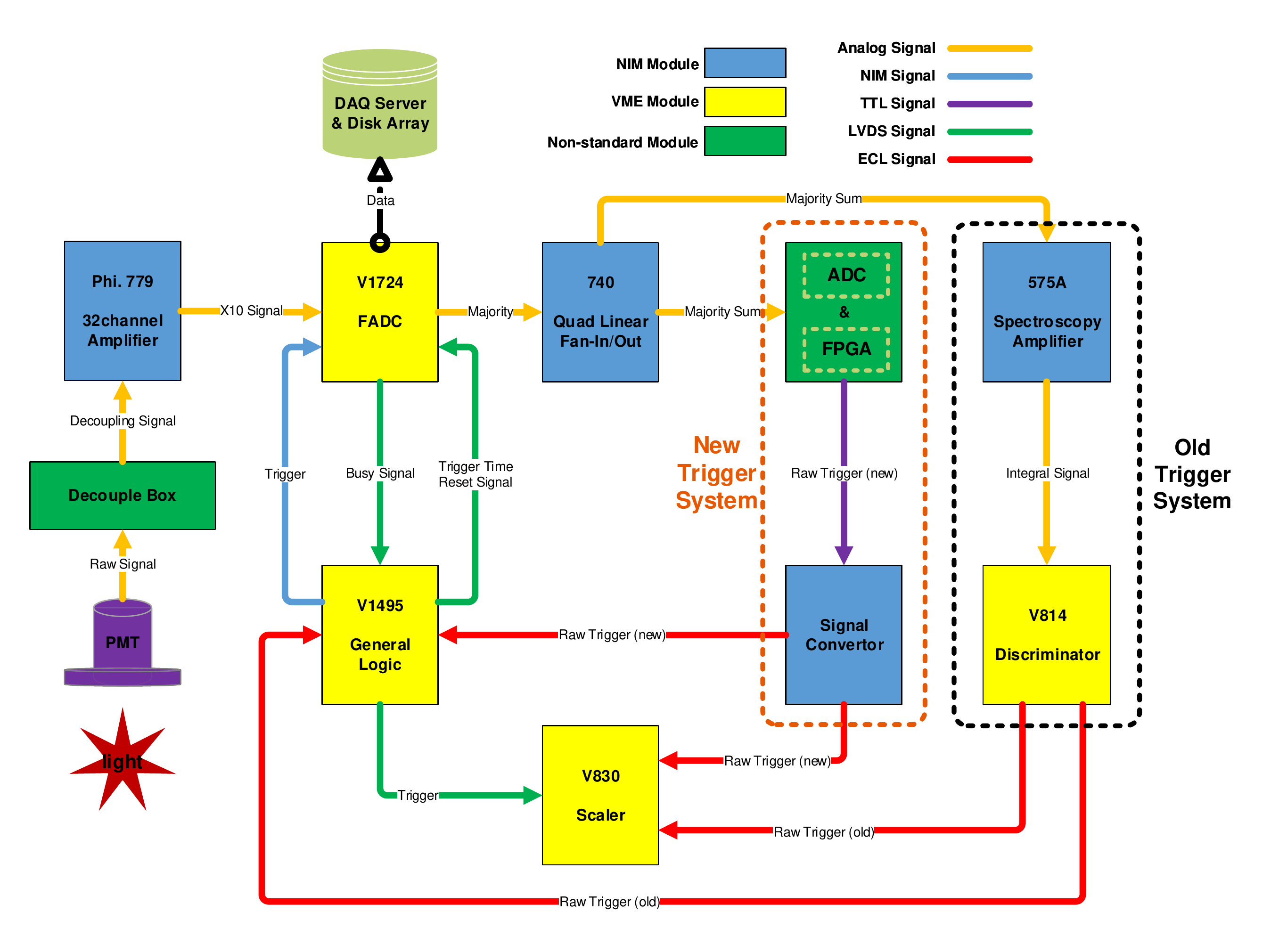}
  \caption{ Diagram of the PandaX-II electronics and DAQ system. The
black dashed lines enclose the old trigger system and the orange
dashed lines enclose the new trigger system.}
  \label{fig:DAQ}
\end{figure}

The data acquisition (DAQ) system of PandaX-II (see Fig.~\ref{fig:DAQ})
is very similar as that from the PandaX-I experiment \cite{PandaX-I}. Signals from
PMTs are amplified by a factor of 10 by Phillips 779 amplifiers, and
then fed into CAEN V1724 8-channel digitizers \cite{1724}. Each channel of the
digitizer can produce a time-over-threshold signal and the internal
sum of these signals from all channels is output as the so-called
Majority (MAJ) signal. MAJ signals from all digitizers and summed by
Phillips 740 linear logic Fan In/Out modules. In the old trigger
system, which was used in the initial phase of data taking, this MAJ sum 
signal was integrated by an ORTEC 575A spectroscopy amplifier. The integrated 
signal was then discriminated by
a CAEN V814 VME discriminator to generate the raw trigger signal. The
raw trigger signal was fed into CAEN V1495 generic logic that generated
the final trigger signal for each V1724 digitizer if not vetoed by the
``BUSY'' signal from any digitizer. A CAEN V830 scaler counted both
the raw and system trigger signals to monitor the system dead-time. More
detailed descriptions of the PandaX-I DAQ can be found in
Ref.~\cite{EnDAQ}.

The old trigger system was relatively simple, but limited in trigger
performance and functionality. It did not analyze detailed
information about the MAJ sum signal, which not only can be used to
generate the trigger signal with lower threshold, but also to veto nonphysical signals
such as spurious waveforms corresponding to discharge process to mitigate the load of the
data acquisition system.

In this paper we describe a new trigger system developed for PandaX-II
experiment.  Instead of integrating the MAJ sum signal, this system
digitizes it, analyzes the digitized data in an FPGA and generates raw trigger
signals.  Section~\ref{sec:des} describes this new trigger system in
detail. Section~\ref{sec:res} presents the trigger efficiency measured 
with real data.

\section{The new PandaX-II trigger system}
\label{sec:des}
\subsection{Overview} 

\begin{figure}[h]\centering
  \includegraphics[width=0.85\textwidth]{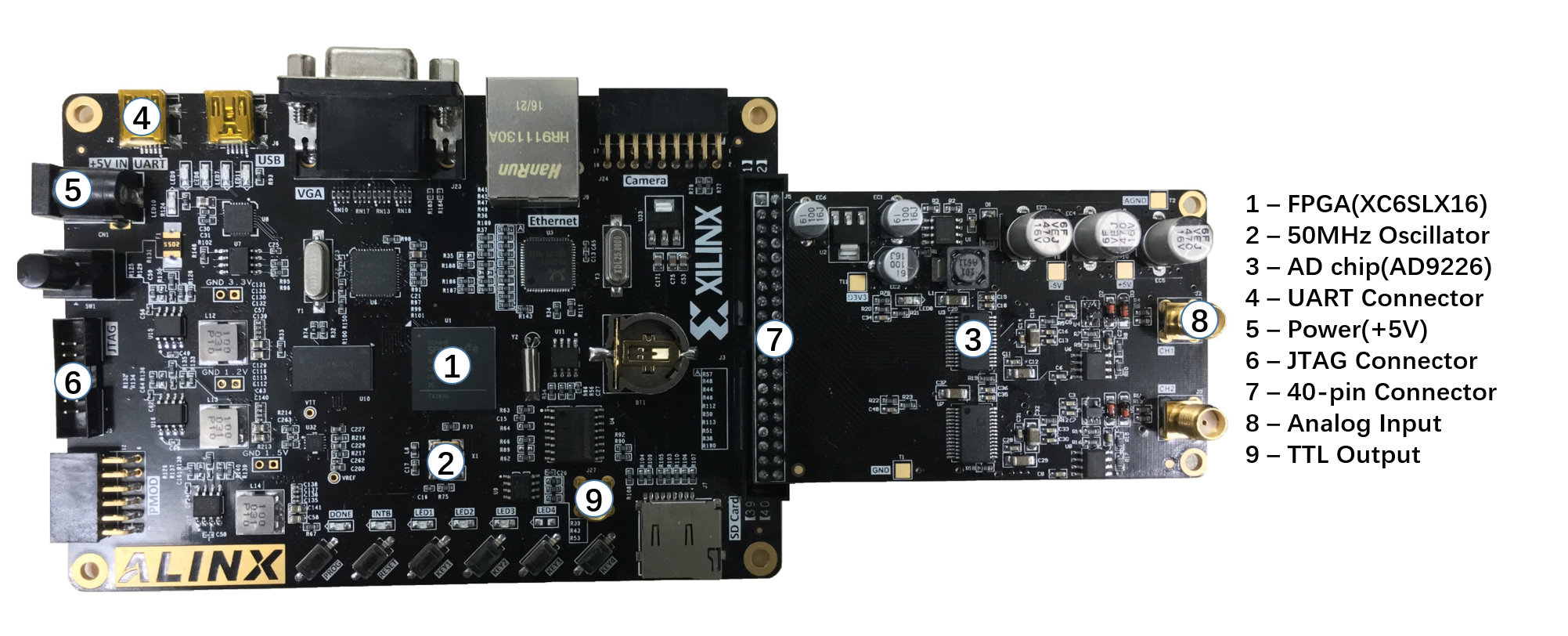}
  \caption{A Photo of the FPGA board and the ADC subboard. The ADC board digitizes MAJ sum signal with 12-bit resolution at 50MHz and sends the data to the Xilinx Spartan-6 XC6SLX16 FPGA. The chip processes the data with the loaded algorithm. Parameters in the algorithm can be configured via the UART port. }
  \label{fig:Board}
\end{figure}

As shown in Fig~\ref{fig:DAQ},The new trigger system mainly consists of an
ADC subboard and an FPGA motherboard (the picture of the two boards are shown in
Fig~\ref{fig:Board}).  The MAJ sum signal is digitized by an AD chip \cite{AD}
(AD9226) with 12-bit resolution at 50 MS/s. The digitized data are then
fed into the FPGA chip (Xilinx Spartan-6). The algorithm implemented in the
FPGA analyzes incoming digitized data and generates a trigger signal once
all trigger conditions are satisfied. The whole system is operated
under the 50 MHz clock provided by the oscillator on the
motherboard. The motherboard output a TTL trigger which is converted
to an NIM signal by a gate generator and then converted to an ECL
signal by an NIM-ECL convertor. As in the old system, converted raw
trigger signals are sent to the generic logic board V1495 for final
trigger decision and to the scaler V830 for monitoring. 


\subsection{Trigger algorithm}

In this section we describe in detail the algorithm implemented in the
FPGA to generate the trigger. The main algorithm calculates the number
of peaks, the accumulated time-over-threshold, and the peak amplitude
of the incoming digitized data and compare them with preconfigured
thresholds. To realize this, we implemented a finite state machine
(FSM) in which the present status only depends on the previous status
and the present data.

\begin{figure}[h]\centering
  \includegraphics[width=0.95\textwidth]{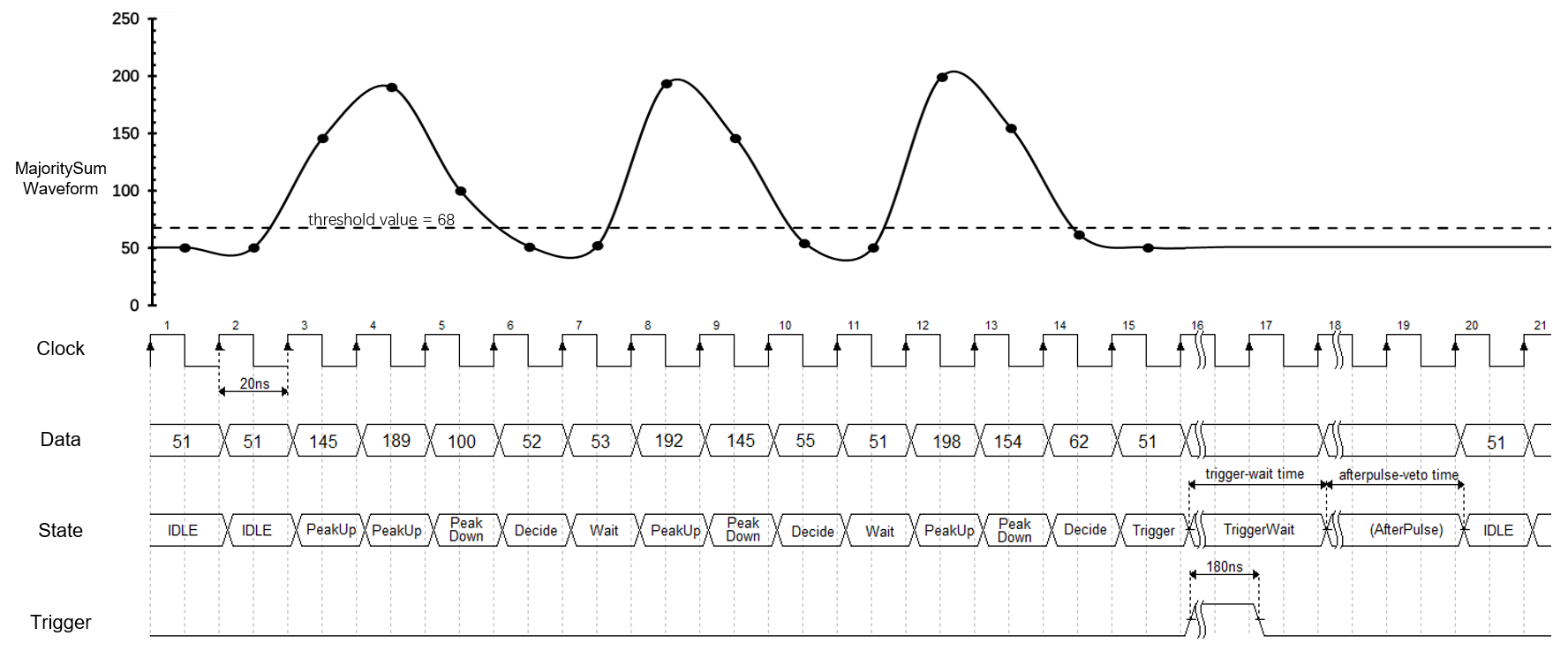}
  \caption{An example of the finite state machine loop in the main algorithm.}
  \label{fig:TimeSequence}
\end{figure}

The main algorithm is illustrated in Fig.~\ref{fig:TimeSequence}. When
the input digitized sample is around the baseline (51 ADC counts in the example),
the FSM will stay in the ``IDLE'' state. When an incoming sample is
larger than the threshold (68 in the example), the FSM will change
to ``PeakUp'' state that indicates an incoming peak and this sample is 
set as the peak amplitude. If the next sample is increasing the FSM will stay in the ``PeakUp'' state
and the peak amplitude will be updated accordingly. If the next sample
is decreasing but larger than the threshold, the FSM will change
to the ``PeakDown'' state that indicates the signal is going back to baseline. 
This peak-finding algorithm will continue untill the sample is below the threshold, when the FSM changes to
the ``Decide'' state to check if all conditions on the above-mentioned
three variables are satisfied. If not satisfied, the FSM will be
in a ``Wait'' status for a predefined time (peak-wait time, several hundred ns) 
until another above-threshold sample appears which indicates an incoming peak again. This is
because a typical MAJ sum signal from S2 contains a few peaks
separated by a few hundred ns.

\begin{figure}[h] \centering
  \includegraphics[width=0.45\textwidth]{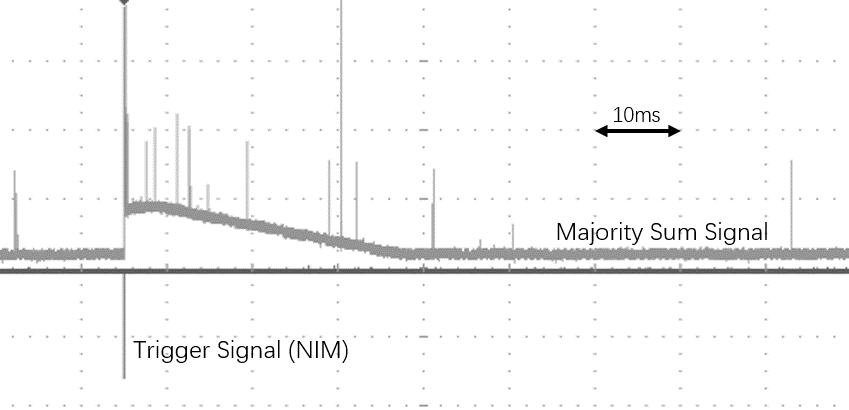}
  \includegraphics[width=0.45\textwidth]{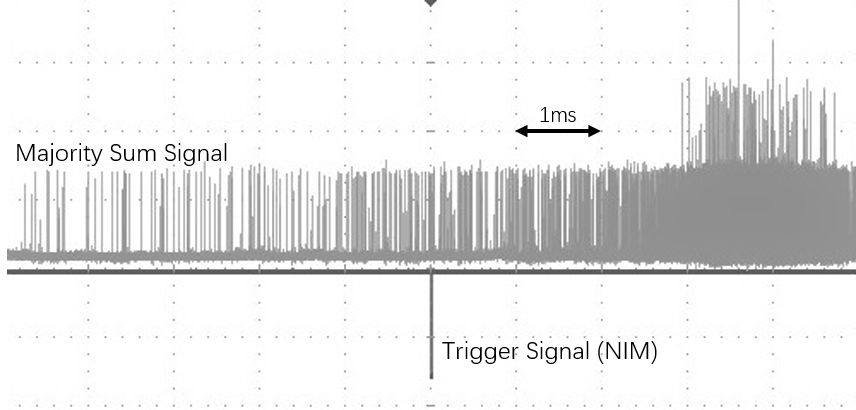}
  \caption{ a) Left: An oscilloscope screen shot of a long afterpulse
following a large S2 signal. The jump of the baseline is due to the Fan-In/Out module that
sums up the majority signals. The tail is larger than 30 ms. The FPGA program
will wait till the afterpulse comes back to baseline and generate
another new trigger when satisfied.  b) Right: An oscilloscope screen shot of a
discharge event. A veto module is implemented in the FPGA to avoid generating 
multiple triggers.}
  \label{fig:ScreenShot}
\end{figure}

Once a trigger signal is generated, the FSM will stay in a
``TriggerWait'' state for a predefined time (trigger-wait time, several ms) to avoid
generating multiple triggers for large S2 signals. For very large S2
signals (indicated by very large peak amplitude) it will wait even
longer till the afterpulse comes back to baseline. After that, the
FSM goes back to ``IDLE'' state and is ready for processing data
and generating a new trigger when all conditions satisfied again.  An
example of a large S2 signal with long afterpulse is shown in
Fig.~\ref{fig:ScreenShot} left, where only one trigger signal is
generated.  We found in the same neutron calibration run condition, 
though the trigger rate of the new trigger system was 30\% higher 
compared to the old one, the raw data size was reduced by 20\% due to 
less afterpulse data were collected. 


Another problem for data acquisition system is the so-called discharge events which
could generate multiple triggers and causes system dead time. The data
exhibit very low amplitude but prolonged pulse trains.  To mitigate
this problem, we implemented a discharge-veto module in FPGA to
identify discharge signals, which usually have quite long accumulated
time-over-threshold. Once a discharge event is identified, a long veto
signal will be applied and no trigger signal will be generated.  An
example of discharge event is shown in Fig.~\ref{fig:ScreenShot}
right. We found in this case, the new trigger system
generated 8 0\% less trigger signals compared to old one.

All the above-mentioned parameters such as peak-wait time, trigger-wait 
time and thresholds are configurable through the UART port shown in 
Fig.~\ref{fig:Board}.

\section{Threshold of the new trigger system}
\label{sec:res}

Besides the above-mentioned ability of the new trigger system to
mitigate problems due to afterpulses or discharging events, another key
performance is the trigger efficiency on S2 signals. To measure the
trigger efficiency, we use after-trigger-S2 signals from events taken in
calibration runs. These S2 signals are recorded in data without
trigger threshold effect. To achieve this, we set
up another system with identical hardware as the new trigger
system. Both systems use the same MAJ sum signal as the input.  One
system is used to generate the trigger for data taking, while another
system, running the same trigger algorithm except that the requirement
on trigger-wait time is relaxed, can still generate trigger output for
these after-trigger-S2s during the same event.  However, these trigger
outputs do not affect the data taking and are only recorded by a V1724
digitizer for the effciency measurement.

\begin{figure}[h]\centering
  \includegraphics[width=0.45\textwidth]{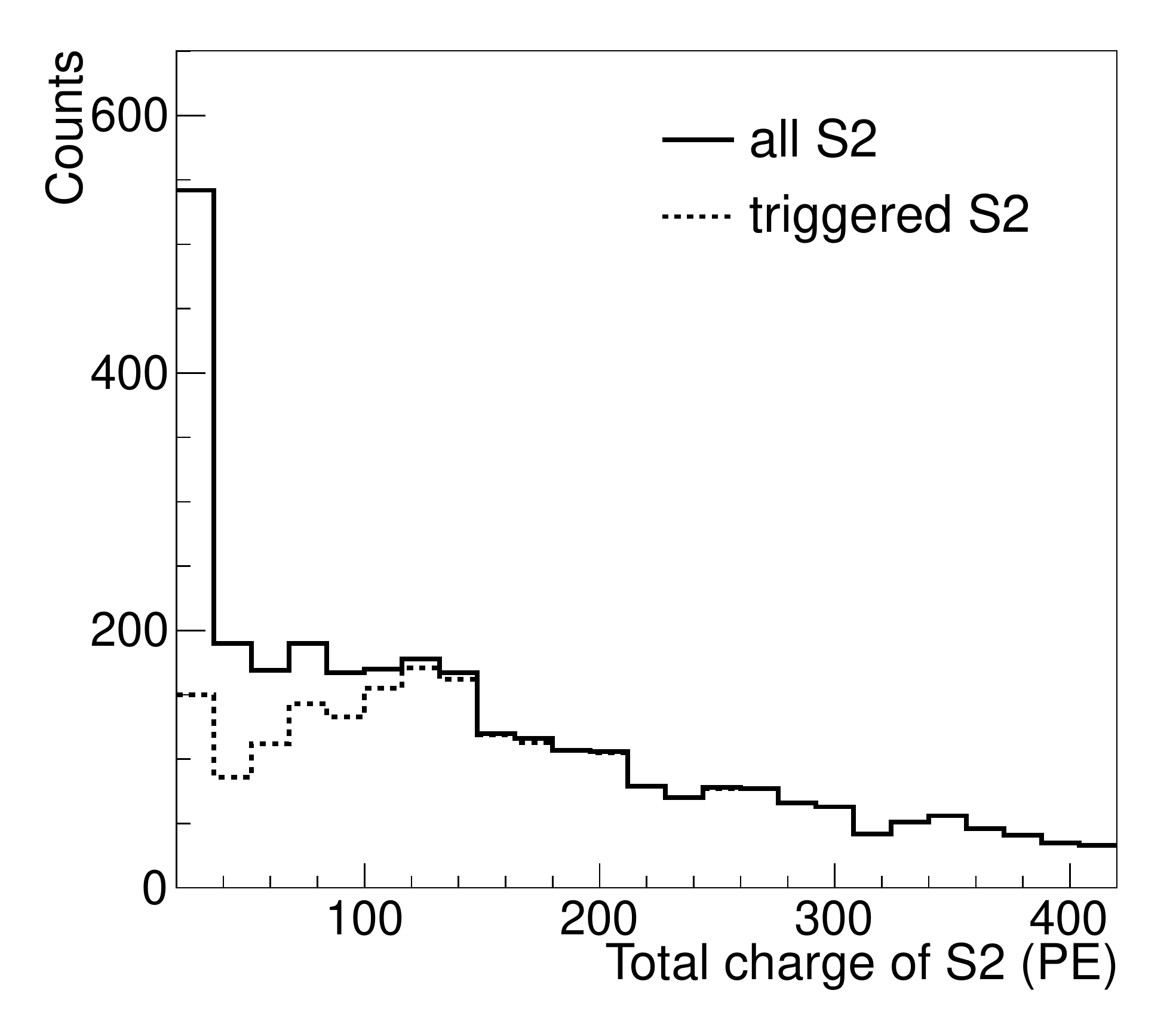}
  \includegraphics[width=0.45\textwidth]{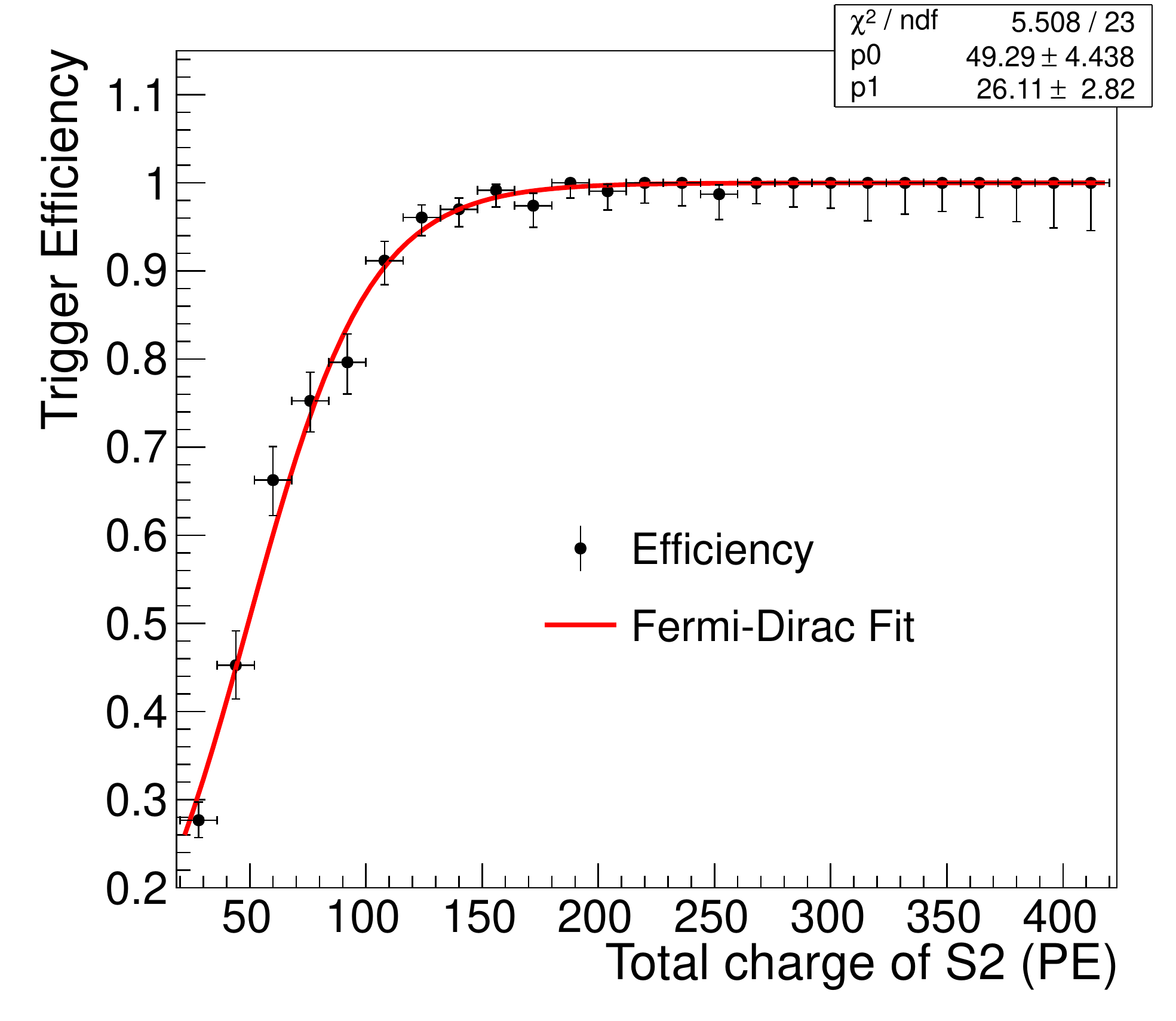}
  \caption{Left: Charge spectrum of all after-trigger-S2 and these
satisfying the trigger condition. Right: trigger efficiency as a
function of the S2 charge.  A fitted Fermi-Dirac function is
superimposed.}
  \label{fig:off_window_s2}
\end{figure}

From the after-trigger-S2 signals and those satisfying the trigger
condition (the spectrum is shown in Fig.~\ref{fig:off_window_s2}
left), we derive the trigger efficiency of the new trigger system,
shown in Fig.~\ref{fig:off_window_s2} right. A Fermi-Dirac ( $f(x) =
1/(e^{\frac{p0-x}{p1}}+1)$ ) fit shows a 50\% trigger efficiency
is achieved for S2 around 50 PE, compared to 79 PE of the old 
trigger system~\cite{PandaX-II_100}. Given that the single electron gain is 
approximately 25 PE/e~\cite{PandaX-II_100}, this corresponds to an average 
trigger threshold of about 2 single electrons.

\section{Summary}

In conclusion, we presented a new trigger system developed for
PandaX-II experiment. Compared to the previous analog-based trigger
system, the new one uses a FPGA device to process digitized data and is
significantly improved on flexibility and performance.  The new system
can effectively reduce mistriggering on afterpulses or discharge events. 
Besides, S2 signals can be triggered with a lower threshold. The new trigger
system has been fully commissioned and deployed for data taking in
PandaX-II.


\acknowledgments
This project has been supported by a 985-III grant from Shanghai Jiao
Tong University, grants from National Science Foundation of China
(Nos. 11435008, 11455001, 11525522), a grant from the Ministry of
Science and Technology of China (No. 2016YFA0400301) and a grant
from the Office of Science and Technology in Shanghai Municipal
Government (No. 11DZ2260700). Yong Yang is partially supported by a grant
from Young 1000-plan program of China (No. GKKQ0720033). We thank Jun Hu and Xiaoshan Jiang
at the Institute of High Energy Physics (IHEP) of the 
Chinese Academy of Sciences for useful discussions and help on 
the trigger algorithm. 



%
%
%
%
%
%

\bibliographystyle{unsrt}
\bibliography{paper}

\begin{thebibliography}{1}

\bibitem{PandaX-II_com}
Andi Tan, Xiang Xiao, Xiangyi Cui, Xun Chen, Yunhua Chen, Deqing Fang, Changbo
  Fu, Karl Giboni, Franco Giuliani, Haowei Gong, et~al.
\newblock Dark matter search results from the commissioning run of
  \uppercase{P}anda\uppercase{X}-\uppercase{II}.
\newblock {\em Physical Review D}, 93(12):122009, 2016.

\bibitem{PandaX-II_100}
Andi Tan, Mengjiao Xiao, Xiangyi Cui, Xun Chen, Yunhua Chen, Deqing Fang,
  Changbo Fu, Karl Giboni, Franco Giuliani, Haowei Gong, et~al.
\newblock Dark matter results from first 98.7 days of data from the
  \uppercase{P}anda\uppercase{X}-\uppercase{II} experiment.
\newblock {\em Physical Review Letters}, 117(12):121303, 2016.

\bibitem{PandaX-II_100SD}
Changbo Fu, Xiangyi Cui, Xiaopeng Zhou, Xun Chen, Yunhua Chen, Deqing Fang,
Karl Giboni, Franco Giuliani, Ke Han, Xingtao Huang, et~al.
\newblock Spin-Dependent WIMP-Nucleon Cross Section Limits from First Data of 
  \uppercase{P}anda\uppercase{X}-\uppercase{II} experiment.
\newblock {\em Physical Review Letters}, 118(7):071301, 2017.

\bibitem{PandaX-I}
Mengjiao Xiao, Xiang Xiao, Li Zhao, Xiguang Cao, Xun Chen, Yunhua Chen, Xiangyi
  Cui, Deqing Fang, Changbo Fu, Karl Giboni, et~al.
\newblock First dark matter search results from the
  \uppercase{P}anda\uppercase{X}-\uppercase{I} experiment.
\newblock {\em Science China Physics, Mechanics \& Astronomy},
  57(11):2024--2030, 2014.

\bibitem{1724}
\uppercase{CAEN V1724} manual.
\newblock {\em http://www.caen.it/csite/}.

\bibitem{EnDAQ}
Xiangxiang Ren, Xun Chen, Xiangdong Ji, Shaoli Li, Siao Lei, Jianglai Liu, Meng
  Wang, Mengjiao Xiao, Pengwei Xie, and Binbin Yan.
\newblock The electronics and data acquisition system for the
  \uppercase{P}anda\uppercase{X}-\uppercase{I} dark matter experiment.
\newblock {\em Journal of Instrumentation}, 11(04):T04002, 2016.

\bibitem{AD}
\uppercase{ANALOG DEVICES}.
\newblock {\em http://www.analog.com/en/index.html}.

\end{thebibliography}
\end{document}